\documentclass{nature}
\usepackage{lineno}
\usepackage{lmodern}
\usepackage{amsmath}
\usepackage{amssymb}
\usepackage{graphicx}
\usepackage{hyperref}
\usepackage{xspace}
\usepackage{caption}
\usepackage{mathptmx}
\usepackage{mathrsfs}
\usepackage{pdfpages}
\usepackage{mathtools} 
\usepackage{bm}

\usepackage{xr-hyper} 
\title{Thermal disorder driven magnetic phases in van der Waals magnet CrI$_{3}$ 
}
\author{Jaume Meseguer-S\'anchez$^1$, Dina Abdul Wahab$^2$, Hubertus Luetkens$^{3}$, Grigol Taniashvili$^4$,
Efr\'en Navarro-Moratalla$^{1, \dagger}$, Zurab Guguchia$^{3,\dagger}$, Elton J. G. Santos$^{5, \dagger}$
}

\makeatletter
\let\saved@includegraphics\includegraphics
\AtBeginDocument{\let\includegraphics\saved@includegraphics}
\renewenvironment*{figure}{\@float{figure}}{\end@float}
\makeatother

\begin{document}

\maketitle

\begin{affiliations}
\item Instituto de Ciencia Molecular, Universitat de Val\'encia, Calle Catedr\'atico Jos\'e Beltr\'an Mart\'inez 2, 46980,  Paterna, Spain 
\item School of Mathematics and Physics, Queen's University Belfast, BT7 1NN, United Kingdom
\item Laboratory for Muon Spin Spectroscopy, Paul Scherrer Institute, CH-5232 Villigen PSI, Switzerland 
\item Department of Physics, Tbilisi State University, Chavchavadze 3, GE-0128 Tbilisi, Georgia
\item Institute for Condensed Matter Physics and Complex Systems, School of Physics and Astronomy, The University of Edinburgh, EH9 3FD, UK  \\
$^{\dagger}$Correspondences to: efren.navarro@me.com, zurab.guguchia@psi.ch, esantos@ed.ac.uk  
\end{affiliations}

\date{}

\begin{abstract}
Magnetic phase transitions often occur spontaneously at 
specific critical temperatures and are instrumental to 
understand the origin of long-range spin order in 
condensed matter systems. The presence of 
more than one critical temperature (T$_{\rm c}$) 
has been observed in several 
compounds\cite{Klauss08,Hiraishi:2014aa,Guguchia:2020aa,Guguchiaeaat3672,Budnick98} 
where the coexistence of competing magnetic orders 
highlights the importance of phase separation driven by 
different factors such as pressure, temperature and chemical composition. 
However, it is unknown whether recently 
discovered two-dimensional (2D) van der Walls (vdW) 
magnetic materials\cite{Huang2017,CrGeTe} 
show such intriguing phenomena that can result in 
rich phase diagrams with novel magnetic features to be explored. 
Here we show the existence of three magnetic phase 
transitions at different T$_{\rm c}$'s  
in 2D vdW magnet CrI$_3$ revealed by 
a complementary suite of muon spin relaxation-rotation ($\mu$SR), 
superconducting quantum interference device 
(SQUID) magnetometry, and large-scale micromagnetic simulations
including higher-order exchange interactions and dipolar fields\cite{Kartsev:2020aa}. 
We find that the traditionally identified Curie 
temperature of bulk CrI$_3$ at 61 K\cite{McGuire2015-dv} 
does not correspond to the long-range order in 
the full volume (V$_{\rm M}$) of the crystal but rather a partial 
transition with less than $\sim$25\% of $V_{\rm M}$ being magnetically spin-ordered. 
This transition is composed of highly-disordered domains with the easy-axis 
component of the magnetization (S$_z$) not being fully spin-polarized 
but disordered by in-plane components (S$_x$, S$_y$) over the entire layer. 
As the system cools down, two additional phase 
transitions at 50 K and 25 K drive the system to 
80\% and nearly 100\% of the magnetically ordered volume, respectively, 
where the ferromagnetic ground state has a marked 
S$_z$ character yet also displaying finite contributions 
of S$_x$ and S$_y$ to the total magnetization.  
Our results indicate that volume-wise competing electronic 
phases play an important role in the 
magnetic properties of CrI$_3$ which set a much lower 
threshold temperature for exploitation in 
magnetic device-platforms than initially considered. 
\end{abstract}

\noindent 


Competing electronic phases underlie a number of unusual physical phenomena 
in condensed matter\cite{Chou06,Uemura2007aa,Dai2008}. From superconductivity up to ferromagnetism, 
when the competition is sizeable the conventional outcome is phase separation. 
Compounds that have shown such behaviour are mostly of complex magnetic 
structures including cuprates\cite{Chou06}, 
iron-based superconductors\cite{Dai2008}, 
ruthenates\cite{Uemura2007aa}, 
topological kagome magnets\cite{Guguchia:2020aa} and 
manganites\cite{Littlewood05,Cheong99}. 
A contrasting case is found in the layered transition 
metal halides\cite{deJonghbook} 
where the presence of 
heavy halide atoms like in CrI$_{3}$ stabilises  
pronounced anisotropy constants 
resulting in long-range magnetic order in 
what appears to be a single ferromagnetic 
transition at a relatively high temperature\cite{Huang2017,Dillon1965-sb}.  
Nevertheless, recent 
experiments\cite{Song:2019aa,Li:2019aa,Liu2018-vc,Wang2018-hq} have unveiled the 
presence of many subtleties in the magnetism of this compound 
which a single transition fails to capture.  
Firstly, CrI$_{3}$ 
exhibits both antiferromagnetic 
and ferromagnetic orders in thin layers
driven by hydrostatic pressure\cite{Song:2019aa,Li:2019aa}.
These phases occur at the same critical temperature 
with a spatial separation of a few hundreds of nanometers
and consequently there is no prelude of thermally-activated spin ordering. 
Secondly, multiple anomalies can be observed in the temperature dependence of the 
magnetic susceptibility below T$_{\rm c}$\cite{Liu2018-vc,Wang2018-hq,McGuire2015-dv}. Such anomalies imply that a more complex magnetic ordering 
involving spins not directly aligned with the easy-axis can emerge. 
Whether different magnetic phases may exist or competition occurs between them is 
largely unknown. However, these observations establish a much more 
intricate scenario than originally pictured for CrI$_{3}$ with 
many hidden features that have important implications 
in the ordering of the magnetic domains in the system. 
Here we use high-resolution ${\mu}$SR spectroscopy, 
complemented by SQUID magnometry and large scale 
micromagnetic simulations, to systematically 
study the thermal evolution of magnetic states in CrI$_3$.
Such suite is instrumental to identify, characterize and understand 
distinct macroscopic ground states with any competing magnetic phases.

%

In a ${\mu}$SR experiment, positive muons implanted into a sample serve 
as extremely sensitive local microscopic probes to detect small internal 
magnetic fields and ordered magnetic volume fractions in the bulk of 
magnetic systems. See details in {\it Methods} and Supplementary Sections S1-S2. 
Zero-field ${\mu}$SR time-spectra are recorded in a powder sample of CrI$_{3}$ below 
(5 K, 30 K, 54 K and 60 K) and above (65 K and 80 K) the magnetic ordering temperature 
(Fig. \ref{fig1}{\bf a-b}). A paramagnetic state is generally characterised by a small 
Gaussian Kubo-Toyabe depolarization of the muon spin originating from 
the interaction with randomly oriented nuclear magnetic moments. 
Conversely, the spectra from the highest measured temperature 
from 150 K down to 62 K, exhibit a relatively high transverse depolarization 
rate ${\lambda}_{T}$ ${\simeq}$ 4.9(2) ${\mu}s^{-1}$. 
This reflects the occurrence of dense electronic Cr 
moments and indicates strong interactions between them.  
In this scenario a novel correlated 
paramagnetic state may be present in the system at 
temperatures above the actual Curie temperature.
As the crystal is cooled down, in addition to the paramagnetic signal, 
an oscillating component with a single well defined frequency 
is observed  at $T \lessapprox 62$ K (Fig. \ref{fig1}{\bf a-b}). 
Below 50 K, a spontaneous muon spin 
precession with two well-separated distinct precession frequencies 
is observed in the ${\mu}$SR spectra and persists down to 5 K. 
The temperature dependences of the internal 
fields (${\mu}_{0}$$H_{\mu}$ = ${\omega}$/${\gamma}_{\mu}^{-1}$) 
for the two components are shown in Fig. \ref{fig2}{\bf a}. 
The low frequency component shows a monotonous decrease with 
increasing temperature and disappears at $T_{\rm {C2}}$ ${\simeq}$ 50 K. 
The high frequency component decreases down to 50 K, 
above which it keeps a constant value within a few Kelvin's 
range and then decreases again to disappear at  $T_{\rm {C1}}$ ${\simeq}$ 62 K. 
Thus, the two oscillatory components have clearly different transition temperatures. 
This implies the presence of two distinct magnetic transitions in CrI$_ {3}$. 
We also notice that an upturn on 
both ${\mu}_{0}$$H_{\mu,1}$ and ${\mu}_{0}$$H_{\mu,2}$ 
is seen below $T_{\rm {C3}}$ ${\simeq}$ 30 K. 
Moreover, a strongly 
damped component appears below $T_{\rm {C3}}$ 
which is seen as some lost of initial asymmetry of the ZF-${\mu}$SR signal. 
This suggests the presence of another magnetic transition at this temperature. 
The temperature 
dependences of the relative weights of the individual components 
in the total ${\mu}$SR signal are shown in Fig. \ref{fig2}{\bf b}. 
The weight of the high frequency component 
(component I) $\omega_{\rm 1}$ gradually increases 
below $T_{\rm {C1}}$ and reaches maximum 
at $T_{\rm {C2}}$, below which the second 
frequency appears. The third component raises 
below $T_{\rm {C3}}$ ${\simeq}$ 30 K. The components I 
and II share the weight of (30 - 70) ${\%}$ in the temperature 
range between 30 K and 50 K. 
These results portray the existence of a clear 
phase diagram in the temperature domain, 
in excellent agreement with the magnetic singularities 
observed by SQUID measurements performed in a single crystal of CrI$_3$ 
(see Supplementary Figure S2).

Fig. \ref{fig2}{\bf c-d} show the temperature dependences 
of the transverse  ${\lambda}_{T}$ and the longitudinal 
${\lambda}_{L}$ depolarisation rates, respectively, of components 
I and II. The ${\lambda}_{T}$ is a measure of the width of the static 
magnetic field distribution at the muon site, and also reflects 
dynamical effects (spin fluctuations). The ${\lambda}_{L}$ 
is determined by dynamic magnetic fluctuations only. 
For both components, ${\lambda}_{T}$ is higher than ${\lambda}_{L}$ 
in the whole temperature range, indicating that magnetism is 
mostly static in origin. However, ${\lambda}_{L1}$ has a higher overall value than 
${\lambda}_{L2}$, implying that the magnetic order 
with $T_{\rm {C1}}$ ${\simeq}$ 62 K 
contains more dynamics. 
The presence of three transitions are clearly 
substantiated by the anomalies, seen 
in ${\lambda}_{T}$  and ${\lambda}_{L}$ (Fig. \ref{fig2}{\bf c-d}). 
Namely, the ${\lambda}_{T,1}$ starts to increase below 
$T_{\rm {C1}}$ and peaks at $T_{\rm {C2}}$, then decreases 
and tends to saturate. Nevertheless, it increments 
again below $T_{\rm {C3}}$. ${\lambda}_{T,2}$ also exhibits 
an increase below $T_{\rm {C3}}$. 
Similarly, ${\lambda}_{L,1}$ goes to high values for  
$T<T_{\rm {C1}}$, saturates at $T<T_{\rm {C2}}$ 
and then enlarges again for $T<T_{\rm {C3}}$, followed by a peak 
at lower temperature. 
We note that it is not possible to discriminate in 
the analysis the contribution of strongly damped 
components and a high frequency component 
into ${\lambda}_{L1}$ below 30 K and thus 
the peak in ${\lambda}_{L2}$ at low temperatures 
could be due to the contribution from component III. 
%
%
%
Overall, these results point to the complex, unconventional thermal evolution of the magnetic states in CrI$_3$.

The behaviour observed involves a volume-wise 
interplay between various states, providing an important 
constraint on theoretical models. One possible interpretation of 
the data is that below $T_{\rm {C1}}$ there is an 
evolution of the magnetic order in specific volumes of the crystal, 
which coexists with a correlated paramagnetic state. 
The second magnetic order thereby  
occurs within the paramagnetic regions below $T_{\rm {C2}}$. 
This interpretation is supported by the temperature dependent 
measurements of the total magnetic fraction $V_{\rm m}$ (Fig. \ref{fig3}). 
The magnetic fraction $V_{\rm m}$ does not acquire the full 
volume below $T_{\rm {C1}}$ ${\simeq}$ 62 K. Instead, it 
gradually increases below $T_{\rm {C1}}$ and 
reaches ${\simeq}$ 80 ${\%}$ at $T_{\rm {C2}}$ ${\simeq}$ 50 K.  
An additional increase of $V_{\rm m}$ takes place 
below $T_{\rm {C3}}$ ${\simeq}$ 25 K, at which the 
third strongly damped component appears and reaches 
nearly ${\simeq}$ 100 ${\%}$. The volume wise evolution 
of magnetic order across $T_{\rm {C1}}$, $T_{\rm {C2}}$ 
and $T_{\rm {C3}}$ in CrI$_ {3}$ strongly suggests the 
presence of distinct magnetic states in the separate 
volumes of the sample. In addition, we propose that while 
the frequency of the second component disappears above 50 K, 
this component exhibits a precursor correlated state up to 
the highest temperature investigated 150 K. 


To understand the microscopic mechanism of 
these phase transitions, we undertake macroscale spin dynamics which 
incorporated atomistic (several \AA's) and micromagnetic 
($\mu$m-level) underlying details of the magnetic and electronic 
structure of CrI$_3$.  
We modelled the atomistic part using 
strongly correlated density functional theory based on 
Hubbard-$U$ methods\cite{Dudarev98} whereas the micromagnetic interactions 
are described through the Landau-Lifshitz-Gilbert (LLG) equation 
techniques\cite{Nakatani_1989}. We have also taken into account 
dipolar fields and higher-order exchange interactions at the level of 
biquadratic exchange since its sizeable magnitude 
is important in the magnetic features of 
2D magnets\cite{Kartsev:2020aa}. 
These approaches have been shown to be critical 
in the description of the spin properties of 
halide magnets at high-accuracy of critical temperatures, 
magnetic domains, and topological spin 
textures\cite{Wahab20,Augustin20}. See further details in Ref.\cite{Kartsev:2020aa}.
We simulated the zero-field cooling processes for a large square flake of 
bulk CrI$_3$~of dimensions 0.4 $\mu$m $\times$ 0.4 $\mu$m 
with a thickness above 200 nm thickness to avoid any 
coexistence of ferro- and anti-ferromagnetic 
phases in the system\cite{Niu:2020aa}. 
As a descriptor of the spin 
dynamics of the system we utilize the ratio between the number 
of spins along the $z-$direction or easy-axis N(S$_{\rm z})$, 
relative to the total number N(S$_{\rm total})$, that is, N(S$_{\rm z})$/N(S$_{\rm total}$). 
This allows us to access the number of sites with a specific spin-polarization 
into the volume of the material which provides information about the disorder
caused by thermal fluctuations. 

We find that three main phases 
emerge in CrI$_3$ as function 
of the temperature, which we name them as Disordered-I, 
Ordered, and Disordered-II (Figure \ref{fig4}). 
Each phase is labelled according to the degree of deviation 
relative to the easy-axis of the magnetization 
S$_{\rm z}$ (Fig. \ref{fig4}{\bf a}). For instance, the Ordered 
phase is characteristic of magnitudes of $|{\rm S_{\rm z}}|$ with high spin 
polarization along $z$, whereas Disordered-II indicates the opposite. 
The Disordered-I however sets an intermediate phase between both states  
with different amount of disorder (Fig. \ref{fig4}{\bf b}). 
Surprisingly, the number of spins in the volume of CrI$_3$
with $|{\rm S_{\rm z}}|>0.99$, corresponding to the full orientation along the easy-axis, 
is largely a minority (N(S$_{\rm z})/$N(S$_{\rm total})\sim 1.8-3.0\%$) at 
the first (T$_{\rm C1}=62$~K) and second (T$_{\rm C2}=50$~K) 
transitions. 
Most of the sites have a substantial amount of disorder with spins 
not completely following the easy-axis of the CrI$_3$ but rather 
stabilising orientations with in-plane orientations (S$_{\rm x}$ and S$_{\rm y}$).  
Such contributions only become negligible as the 
temperature drops down below 10 K where a sharp increment of 
N(S$_{\rm z})$/N(S$_{\rm total}$) appears which coincides with the 
antagonistic response of the Disordered-I phase. 
Spin sites with larger amount of disorder (e.g. Disordered-II) 
tend to disappear as the system cools down below 40 K. 
A spatial visualization of the spin orientations, shown in Fig. \ref{fig4}{\bf c} 
for different temperatures confirm this description. 
Remarkably, it shows that mutually exclusive spatial 
magnetic regions with distinct projection of the magnetization 
exist in a wide temperature range, which is in perfect 
agreement with ${\mu}$SR results. However, in ${\mu}$SR the 
inhomogeneous and mutually exclusive spatial regions exist 
even at the base-$T$ of 5 K, while the calculations show 
a sharp increment of the 
homogeneous $S_{\rm z}$-axis oriented structure below 10 K 
(Figure \ref{fig4}{\bf b}). 
The reason for this discrepancy could be that: (i) there is not a  
one-to-one correspondence in the temperature values 
between experiment and theory. (ii) In reality, much lower temperatures 
than 5 K are most likely required to get homogeneous 
magnetic states. It is worth mentioning that even when the system 
reached 0 K, where thermal fluctuations are inexistent, the spins still evolve in time as an
effect due to high magnetic anisotropy and meta-stability of the 
magnetic domains in CrI$_3$\cite{Wahab20} (Fig. \ref{fig4}{\bf c}). 
Our results suggest that the magnetic structure of bulk CrI$_3$ is markedly 
composed by thermal disorder with no apparent formation of 
anti-ferromagnetic phases as observed 
in few layers \cite{Huang2017,Niu:2020aa,Song:2019aa,Li:2019aa}.

\section*{Supplementary Materials}
\label{sec:org3881bef}

Materials and Methods. 
\\
Supplementary sections S1 to S5, and Figures S1-S2.

\subsubsection{Data Availability}

The data that support the findings of this study 
are available within the paper and its Supplementary Information.  

\subsubsection{Competing interests}
The Authors declare no conflict of interests.

\subsubsection{Acknowledgments}
${\mu}$SR experiments were performed at 
at the $\pi$M3 beam line (low background GPS  instrument) 
of the Swiss Muon Source (SmuS) of the 
Paul Scherrer Insitute, Villigen, Switzerland, under 
proposal ID: 20190297 with EJGS as the PI. 
GT thank Prof. Alexander Shengelaya and the Georgian National 
Science Foundation (grant PHDF-19-060) for funding support to participate in 
${\mu}$SR experiments led by ZG and EJGS. 
EJGS acknowledges computational resources through the 
UK Materials and Molecular Modelling Hub for access to THOMAS supercluster, 
which is partially funded by EPSRC (EP/P020194/1); CIRRUS Tier-2 HPC 
Service (ec131 Cirrus Project) at EPCC (http://www.cirrus.ac.uk) funded 
by the University of Edinburgh and EPSRC (EP/P020267/1); 
ARCHER UK National Supercomputing Service (http://www.archer.ac.uk) via 
Project d429. EJGS acknowledges the 
EPSRC Early Career Fellowship (EP/T021578/1) and 
the University of Edinburgh for funding support. 
ENM acknowledges the European Research Council 
(ERC) under the Horizon 2020 research and innovation programme 
(ERC StG, grant agreement No. 803092). 
%

\section*{References and Notes} 
\bibliography{references.bib} 

\pagebreak{}


\subsubsection*{Figure captions}


\begin{figure*}[b!]
\caption{{\bf $\mu$SR spectroscopy applied to CrI$_3$.} 
{\bf a-b,} Zero-field ${\mu}$SR spectra, recorded at various temperatures for the polycrystalline sample of CrI$_ {3}$, 
shown in the low and extended time interval. The solid lines are the fit of the data 
using the methods of Supplementary Sections S3-S4. 
Error bars are the standard error of the mean 
in about 10$^6$ events. The error of each bin count 
is given by the standard deviation of $n$. 
The errors of each bin in the $\mu$SR asymmetry 
are then calculated by statistical error propagation. 
}
\label{fig1}
\end{figure*}

\begin{figure*}[b!]
\caption{{\bf Temperature dependent $\mu$SR parameters.} {\bf a,} The temperature dependence of the internal magnetic fields for the observed two components in CrI$_ {3}$. {\bf b,} The temperature dependence of the relative weights of the three components in the total signal for CrI$_ {3}$, determined from zero-field ${\mu}$SR experiments. {\bf c-d,} The temperature dependence of transverse depolarization rates ${\lambda_{T1}}$, ${\lambda_{T2}}$ and the longitudinal depolarization rates ${\lambda_{L1}}$, ${\lambda_{L2}}$ for two components.
The error bars represent the standard deviation of the fit parameters.
}
\label{fig2}
\end{figure*}

\begin{figure*}[b!]
\caption{{\bf Thermal evolution of various magnetic phases in CrI$_3$.} {\bf a,} The temperature dependence 
of the total magnetic volume fraction $V_M$, determined from precise weak transverse field (weak-TF) ${\mu}$SR measurements. In this weak-TF experiment, 
a small magnetic field of 30 G is applied nearly perpendicular to the muon 
spin polarisation. The different components seen in Fig. \ref{fig2} 
are highlighted in each region of the temperature range with 
the paramagnetic phase above the Curie temperature. The error bars represent the standard deviation of the fit parameters. 
}
\label{fig3}
\end{figure*}

\begin{figure*}[b!]
\center
\caption{\label{fig4}\textbf{Micromagnetic analysis of the spin dynamics in CrI$_3$.}
{\bf a,} Diagram of the different magnitudes of $|{\rm S}_{\rm z}|$ relative 
to the unit sphere in the range of $|{\rm S}_{\rm z}|=1$ (full polarized along $z$) 
and $|{\rm S}_{\rm z}|=0$ (only in-plane $xy-$projections). The orientations 
of the spins are assigned randomly following the Monte Carlo moves. 
The plot only shows a schematic of possible projections over the sphere 
with others (not shown) with the same $|{\rm S}_{\rm z}|$ equally probable. 
{\bf b,} Magnetic fraction (in \%) of the out-of-plane 
spins (N(S$_{\rm z})$/N(S$_{\rm total}$)) 
as a function of the temperature. The calculation 
of N(S$_{\rm z})$ and N(S$_{\rm total}$) take into 
account the volumetric amount of spins over the entire system for S$_{\rm z}$ and 
S$_{\rm total}$, respectively. 
Three main phases were identified in bulk CrI$_3$ 
which are named: 
Disordered-I (faint black), with values of N(S$_{\rm z})$/N(S$_{\rm total}$) 
between the curves of $0.01<|{\rm S}_{\rm z}|<0.99$ and $0.20<|{\rm S}_{\rm z}|<0.80$; 
Ordered (faint red), within $|{\rm S}_{\rm z}|>0.80$ and  $|{\rm S}_{\rm z}|>0.99$; and, 
Disordered-II (faint blue), within $|{\rm S}_{\rm z}|<0.01$ and $|{\rm S}_{\rm z}|<0.20$. 
The solid curves showed in each phase provide a sample of a 
specific variation of N(S$_{\rm z})$/N(S$_{\rm total}$) for a given range of 
$|{\rm S}_{\rm z}|$. The vertical dashed line sets the time of 2 ns 
required to achieve 0 K in the spin dynamics. The magnitudes of 
N(S$_{\rm z})$/N(S$_{\rm total}$) showed after 0 K demonstrated 
that even when the thermal fluctuations are zero, the system is still 
evolving to stabilize its ground state. The critical 
temperatures (T$_{\rm C1, C2,C3}$)
are also highlighted. 
{\bf c,} Snapshots of the dynamical spin configurations of 
bulk CrI$_3$ during field cooling from 80 K down to 0 K. 
Magnetisation is projected along of in-plane components 
(S$_{\rm x}$, S$_{\rm y}$) and 
S$_{\rm z}$  at different temperatures: 68 K, 60 K, 50 K, 20 K and 0 K. 
Each column and row corresponds to a specific projection of the magnetization 
at a given temperature provided at the far left. The colour scale shows 
the variation of the three magnetization components (S$_{\rm x}$, S$_{\rm y}$, S$_{\rm z}$)  
throughout the system. 
}  
\label{fig4}
\end{figure*}

\pagebreak{}

\clearpage


\subsubsection*{Figures}

\setcounter{figure}{0}


\begin{figure}[htbp]
\centering
\includegraphics[width=0.99\linewidth]{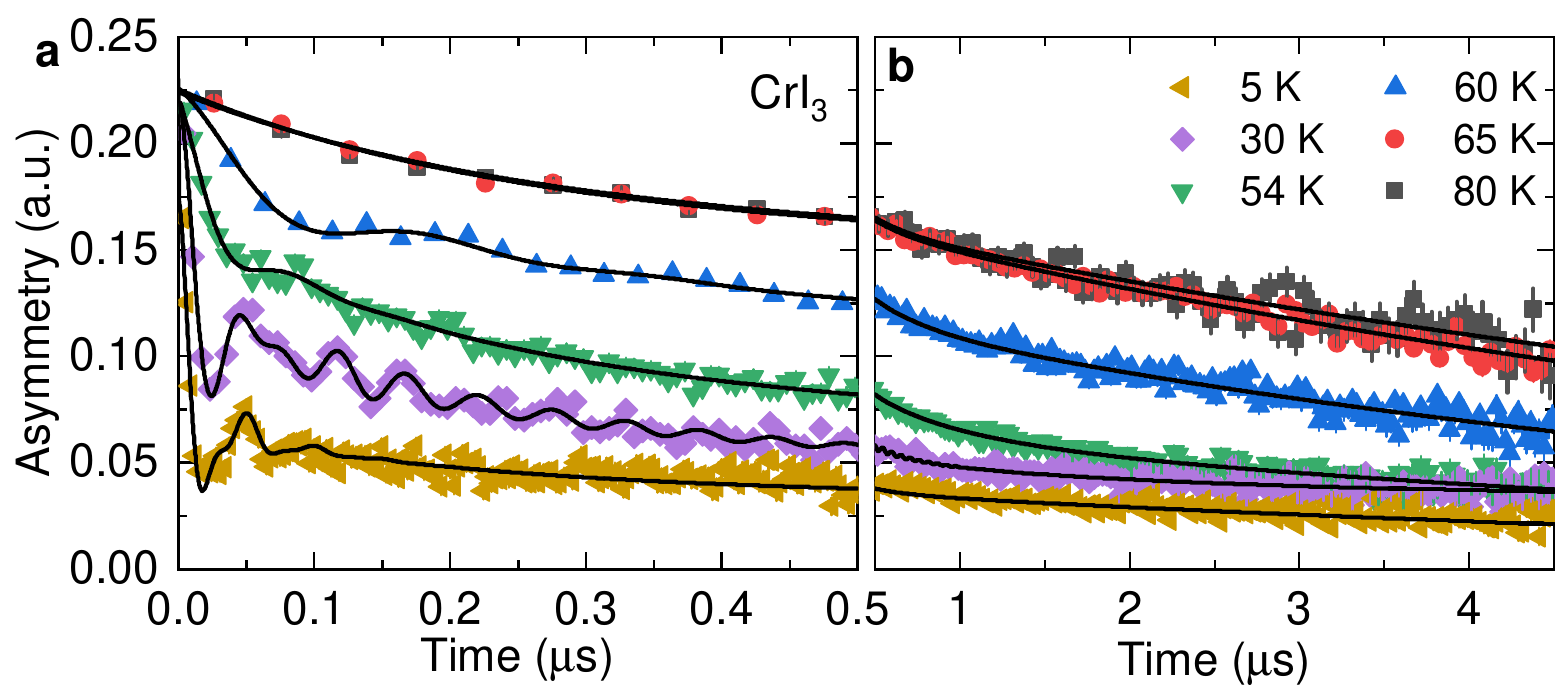} 
\caption{}
\end{figure}

\begin{figure}[htbp]
\centering
\includegraphics[width=0.99\linewidth]{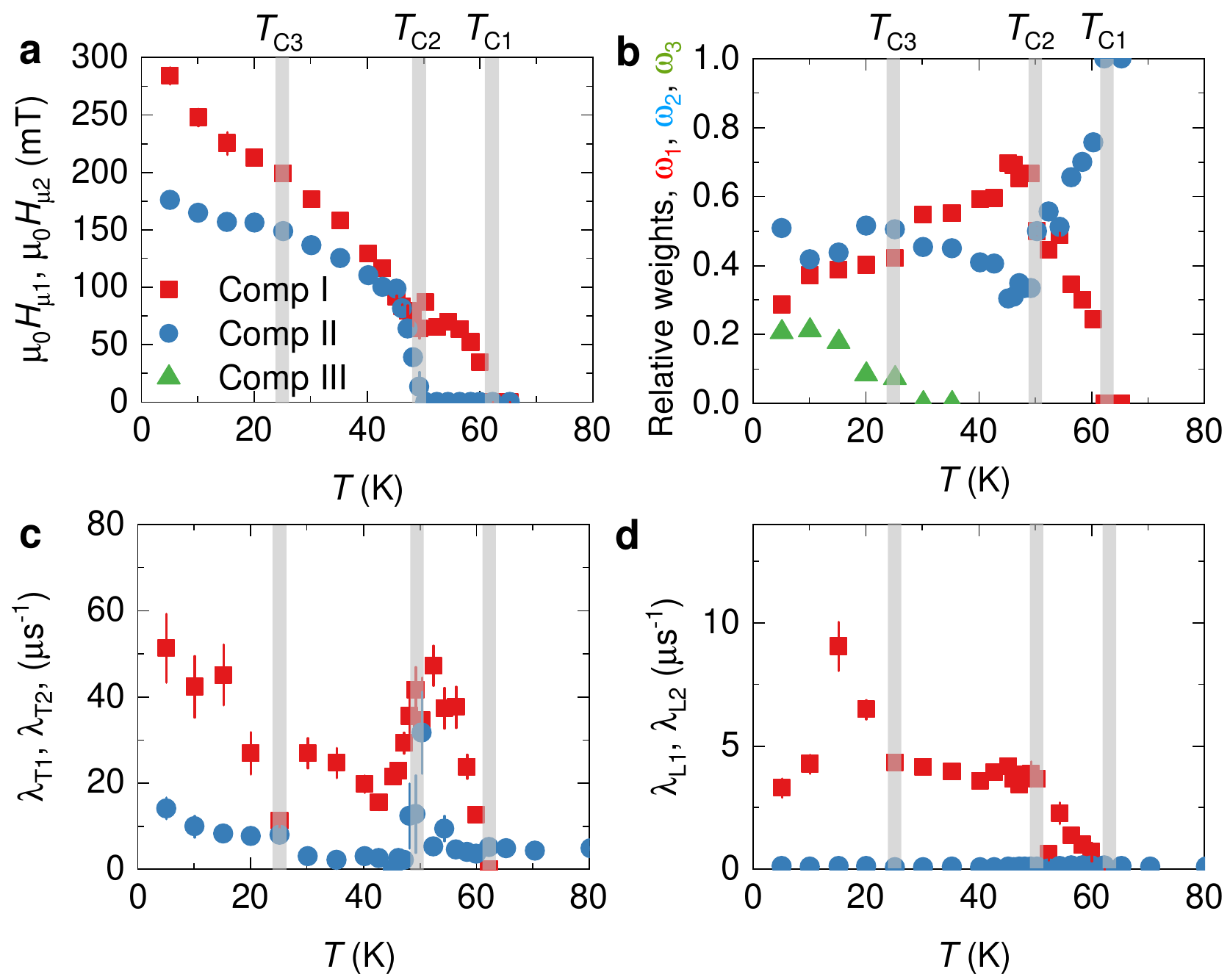} 
\caption{}
\end{figure}

\begin{figure}[htbp]
\centering
\includegraphics[width=0.99\linewidth]{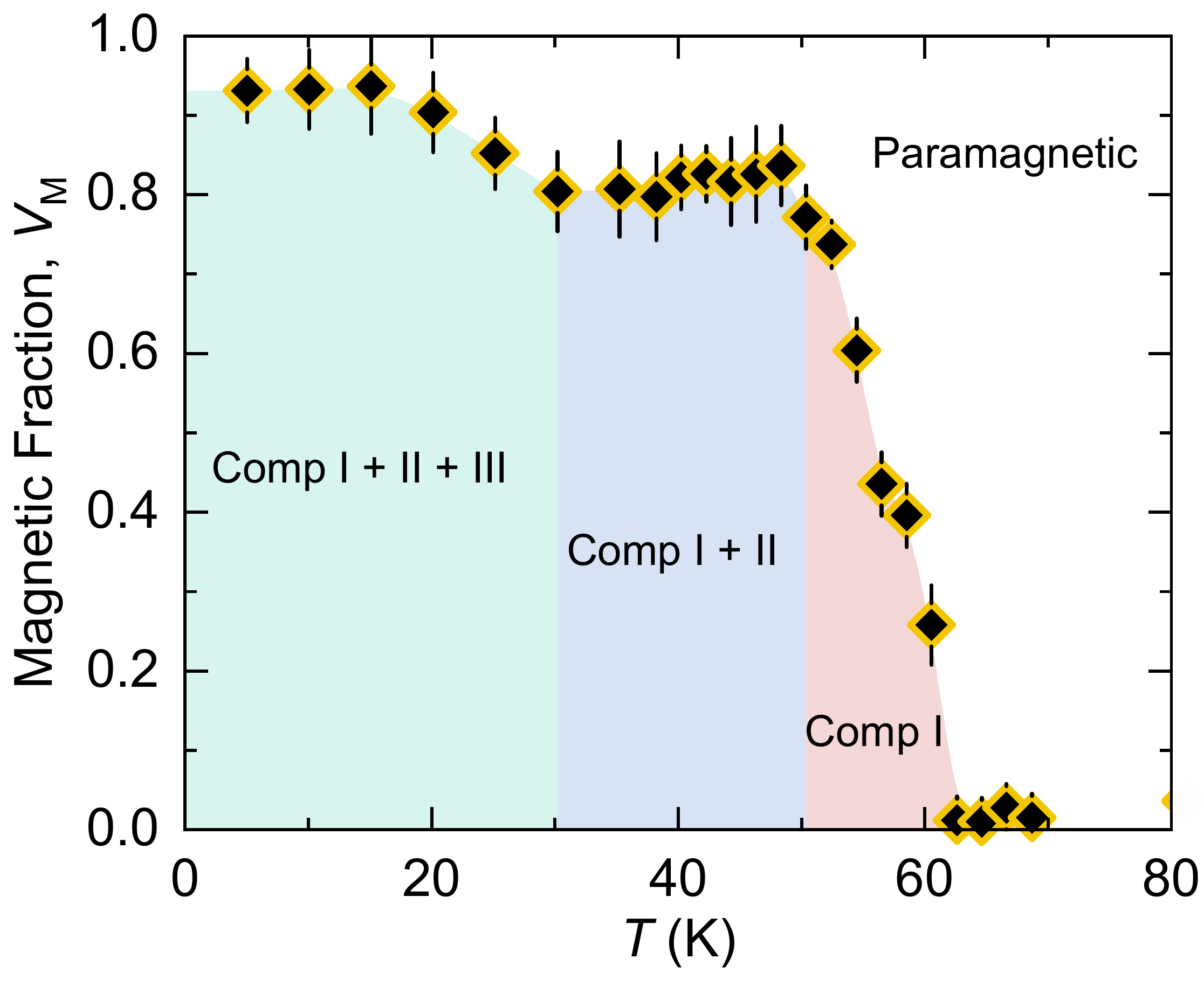} 
\caption{}
\end{figure}

\begin{figure}[htbp]
\centering
\includegraphics[width=1\linewidth]{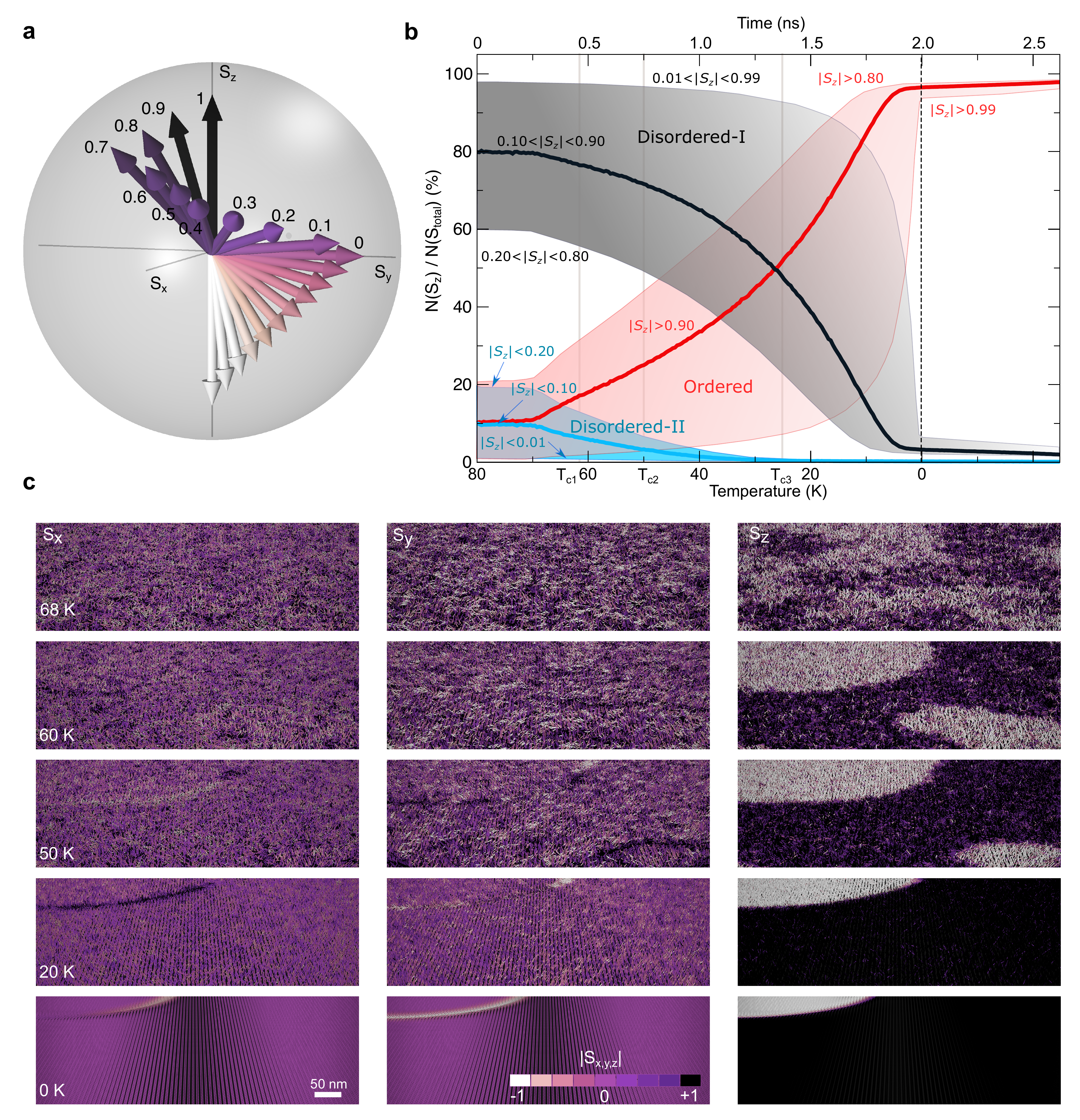} 
\caption{}
\end{figure}

\pagebreak{}


\end{document}